\documentclass[journal=jacsat,manuscript=article]{achemso}

\usepackage[version=3]{mhchem} 
\setkeys{acs}{maxauthors = 0}

\author{Robert J. Kirby}
\affiliation{Department of Chemistry, Princeton University, Princeton, New Jersey 08544, USA}
\author{Austin Ferrenti}
\affiliation{Department of Chemistry, Princeton University, Princeton, New Jersey 08544, USA}
\author{Caroline Weinberg}
\affiliation{Department of Chemistry, Princeton University, Princeton, New Jersey 08544, USA}
\author{Sebastian Klemenz}
\affiliation{Department of Chemistry, Princeton University, Princeton, New Jersey 08544, USA}
\author{Mohamed Oudah}
\affiliation{Department of Chemistry, Princeton University, Princeton, New Jersey 08544, USA}
\alsoaffiliation{Quantum Matter Institute, University of British Columbia, Vancouver, British Columbia, V6T 1Z4, Canada}
\author{Chris P. Weber}
\affiliation{Department of Physics, Santa Clara University, 500 El Camino Real, Santa Clara, California 95053-0315, USA}
\author{Daniele Fausti}
\affiliation{Department of Chemistry, Princeton University, Princeton, New Jersey 08544, USA}
\alsoaffiliation{Department of Physics, Università degli Studi di Trieste, Trieste I-34127, Italy}
\alsoaffiliation{Sincrotrone Trieste S.C.p.A., Basovizza, Trieste I-34012, Italy}
\author{Gregory D. Scholes}
\affiliation{Department of Chemistry, Princeton University, Princeton, New Jersey 08544, USA}
\author{Leslie M. Schoop}
\email{lschoop@princeton.edu}
\affiliation{Department of Chemistry, Princeton University, Princeton, New Jersey 08544, USA}

\title{Transient Drude Response Dominates Near-Infrared Pump-Probe Reflectivity in Nodal-Line Semimetals ZrSiS and ZrSiSe}

\abbreviations{IR,NMR,UV}
\keywords{American Chemical Society, \LaTeX}

\begin{document}

\begin{tocentry}

\includegraphics[width=0.85\textwidth]{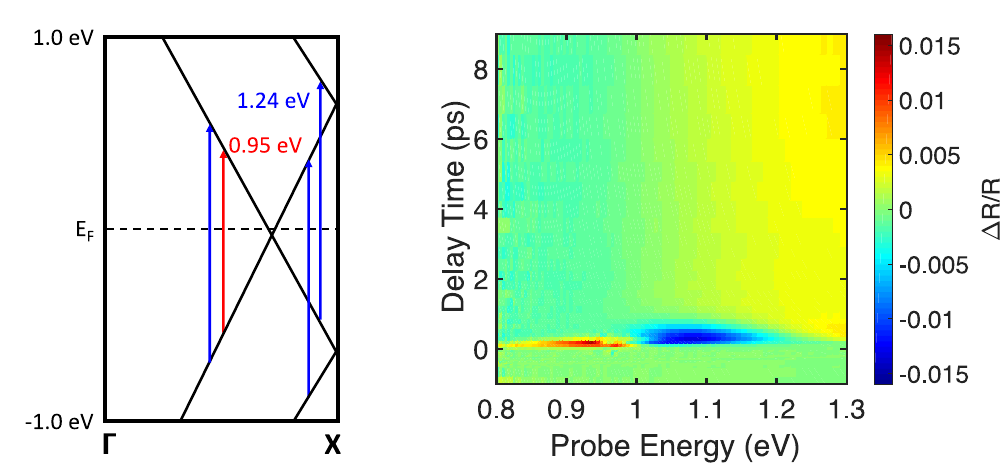}

\end{tocentry}

\begin{abstract}
The ultrafast optical response of two nodal-line semimetals, ZrSiS and ZrSiSe, was studied in the near-infrared using transient reflectivity. The two materials exhibit similar responses, characterized by two features, well resolved in time and energy. The first transient feature decays after a few hundred femtoseconds, while the second lasts for nanoseconds. Using Drude-Lorentz fits of the materials' equilibrium reflectance, we show that the fast response is well-represented by a decrease of the Drude plasma frequency, and the second feature by an increase of the Drude scattering rate. This directly connects the transient data to a physical picture in which carriers, after being excited away from the Fermi energy, return to that vicinity within a few hundred femtoseconds by sharing their excess energy with the phonon bath, resulting in a hot lattice that relaxes only through slow diffusion processes (ns). The emerging picture reveals that the sudden change of the density of carriers at the Fermi level instantaneously modifies the transport properties of the materials on a timescale not compatible with electron phonon thermalization and is largely driven by the reduced density of states at the nodal line.
\end{abstract}

\section{Introduction}

The physical realization of topological semimetals (TSM), starting with the 2-dimensional Dirac semimetal (DSM) graphene \cite{Novoselov2004}, opened the door for researchers to study the myriad of exotic physical phenomena predicted to occur in these materials. TSMs have a characteristic electronic structure, in which bulk linear bands \textemdash\ one from the valence band and one from the conduction band \textemdash\ cross at a point, points, or over a line in $k$-space. One consequence of these linear band crossings is that quasiparticle excitations within them are predicted to be solid-state analogs to high-energy excitations that obey relativistic Hamiltonians, such as the Dirac or Weyl Hamiltonian, which both have a linear energy-momentum dispersion relation. In the solid state, this results in carriers having high mobility \cite{Liang2015}, suppressed back-scattering \cite{Roushan2009}, and short excited-state lifetimes, making TSMs attractive materials for high-frequency optoelectronic devices, including saturable-absorber mirrors \cite{Zhu2017, Meng2018} or ultrafast infrared (IR) photodetectors \cite{Chan2017, Schoop2018}. Of all the different flavors of TSMs, nodal-line semimetals (NLSMs), in which band crossings extend over a line in momentum space, present themselves as superior materials for these applications; the extension of the crossing through the Brillouin zone (BZ) increases the density of states (DOS) of carriers in these bands, providing a greater density of carriers with which photons can interact.
 
Two such promising NLSMs are ZrSiS and ZrSiSe, tetragonal materials in the \textit{P4/nmm} space group that crystallize in the PbFCl structure type \cite{Schoop2016}. The crystal structure is illustrated in Fig.\ 1 \textbf{a}. The band structures of these two materials are defined by their nodal-line crossings, which form a diamond-shaped loop of crossings around the $\Gamma$-point and have bandwidths exceeding 2 eV \cite{Schoop2016, Topp2016}. These bands arise from the stabilizing interaction between the $p_x$- and $p_y$-orbitals in the Si square net and the nearby Zr $d$-orbitals \cite{Schoop2016}. The nodal line itself is very close to the Fermi energy (tens of meV), the bands of which are the only bands in a $\sim1$ eV window straddling $E_F$, making these ideal materials in which to study the linear bands without interference. In addition to the nodal-line structure, the two materials also exhibit Dirac crossings at the $X$-point $\sim0.5$ eV above and below $E_F$ \cite{Topp2016}; these crossings are protected by the non-symmorphic symmetry of the crystal structure. The relative distance of the crossings from the Fermi energy is inversely proportional to the distance between the Si atoms in the Si-Si square net in the crystal structure \cite{Klemenz2020}. The Si-Si distance in ZrSiSe is slightly larger than in ZrSiS, so the Dirac crossings are nearer the Fermi energy ($\pm$ 0.6 eV vs $\pm$ 0.7 eV for ZrSiS), resulting in a slightly squashed band structure when compared with that of ZrSiS. Spin-orbit coupling (SOC) induces a 30 meV gap in the nodal line of ZrSiS \cite{Schill2017}, and should give rise to a slightly larger gap in ZrSiSe, due to the increased atomic weight of the chalcogen.

\begin{figure}
\includegraphics[width=0.5\textwidth]{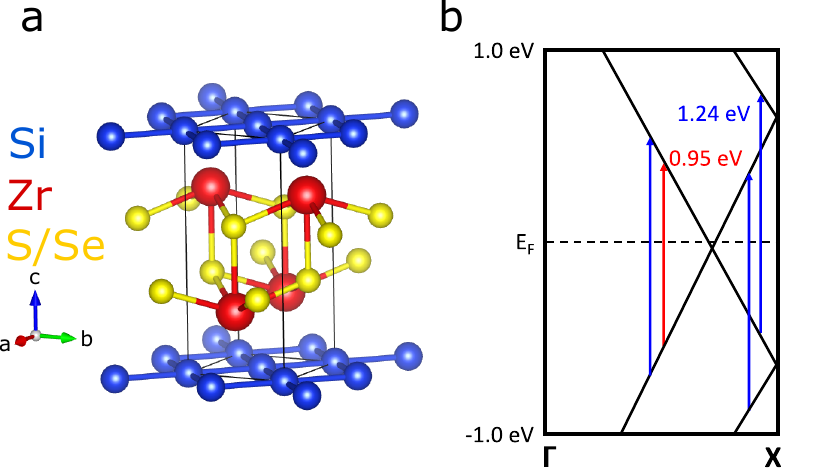}
\caption{\label{fig} \textbf{(a)} Crystal structure of ZrSiS and ZrSiSe.  \textbf{(b)} Schematic of the band structure of ZrSiS between $\Gamma$ and $X$. The nodal line bands reside between $\Gamma$ and $X$, crossing slightly below $E_F$, and the Dirac nodes are at the $X$ point. The two pump energies used in our experiments are shown, as well as the interband transitions they can excite.}
\end{figure}

Optical transitions within the nodal line bands and the other linear bands comprising the Dirac crossings dominate the joint density of states (JDOS) from low photon energies up to $\sim1.5$ eV photon energies, at which point transitions from linear bands to higher-energy trivial bands start to take over \cite{Habe2018, Ebad2019}. Therefore, up until the visible region, these are ideal materials with which to harness the optoelectronic promise of the nodal line. In order to assess these materials' applicability in such devices, however, their electronic response to light on the relevant ultrashort timescales (femtoseconds to nanoseconds) must first be characterized and understood.

The two previous ultrafast optical spectroscopic studies of ZrSiS \cite{Weber2017, Weber2018} were part of attempts to find unifying characteristics in the electronic relaxation in TSMs by comparing the response of different TSMs, from different space groups and structure types, using transient reflectivity (to the best of the authors' knowledge, the ultrafast optical response of ZrSiSe is as yet unexamined). In this technique, a  ``pump" laser pulse optically excites carriers and, after a time delay, a weaker ``probe" pulse interrogates how the reflectivity of the sample changes after photoexcitation. Variation in the delay time between the pulses is then used to track how the photoexcited state evolves and returns to equilibrium. Both previous studies used narrowband pulses: the first in the visible and near-infrared (NIR) \cite{Weber2017}, and the second in the mid-infrared (MIR) \cite{Weber2018}. 

In the first study \cite{Weber2017}, the transient response of ZrSiS could be fit well with a tri-exponential decay with lifetimes of 190 fs, 1.6 ps, and 100 ps. The processes behind these lifetimes were assigned, respectively, to thermalization (the process in which a small, initially non-equilibrium carrier population exchanges energy rapidly within itself to reach a quasi-equilibrium population with a temperature higher than that of the lattice) or cooling of a thermalized excited population (through the emission of high-energy optical phonons), and to bimodal cooling kinetics \cite{Weber2017}. The second transient study \cite{Weber2018} found that the evolution of the photoexcited state in the MIR was best described by a mono-exponential decay with lifetimes increasing from 200 fs to 400 fs as the pump energy was increased from 86 meV to 500 meV. This process was ascribed to carrier cooling (again, through phonon emission), and thermalization was suggested to have occurred within the time resolution of the experiment ($\sim100$ fs) due to the large DOS residing in the nodal line bands \cite{Weber2018}. It has, however, proven difficult to derive any concrete conclusions about which physical process lies behind each lifetime, since these assignments are obtained by analogy to the relaxation dynamics ascertained for other materials which have been more comprehensively studied, such as the DSMs graphene \cite{Johannsen2013, Breusing2011, Brida2013}, and Cd$_3$As$_2$ \cite{Lu2017, Weber2015, CZhu2017}. 

To this end, we investigated how the optical properties (i.e.\ the reflectivity) of the isostructural NLSMs ZrSiS and ZrSiSe respond on the femtosecond to picosecond timescale using transient reflectivity. Our aim in comparing isostructural materials is to isolate the response from the nodal line bands, and remove ambiguity in the dynamics that could arise from comparing materials with different crystal and band structures. We chose to use two pump energies, 0.95 eV and 1.24 eV, to discern if there is a difference between excitations inside and outside the nodal-line bands (Fig.\ 1 \textbf{b}). The calculated optical conductivity for ZrSiS in Ref. \cite{Ebad2019} shows that transitions at 0.95 eV are predominantly between nodal-line bands with a small contribution from nodal line-to-Dirac cone transitions, whereas the optical conductivity at 1.24 eV, while still seeing a contribution from inter-nodal line transitions, contains a greater contribution of nodal line-to-Dirac cone transitions. What separates our work from previous studies is the spectral region in which we are pumping, the NIR, and the broadband nature of the probe pulse (which also covers the NIR). This allows us to better connect the observed out-of-equilibrium spectral features with the materials' equilibrium reflectance. Moreover, through an understanding of the equilibrium reflectance, we elucidate the physical processes behind the transient features; lattice heating, for example, which can be identified through fairly well-defined spectral changes.

\section{Results and discussion}

The pump-probe data for ZrSiS and ZrSiSe, pumped at 1.24 eV and 0.95 eV, with a fluence of 0.68 mJ/cm$^{2}$, can be seen in Fig.\ 2. Starting with ZrSiS pumped at 1.24 eV, Fig.\ 2 \textbf{a}-\textbf{c} tracks the change in the material's reflectivity, $\frac{\Delta R}{R}$, as a function of probe energy and pump-probe delay time. This spectrum consists of three main features, well separated in time and energy: the coherent artefact ($t \sim 0.1$ ps, also called cross-phase modulation), a negative peak at 1.08 eV ($t \sim 0.3 - 1$ ps), and a broad feature that is negative at low probe energies and positive at higher probe energies, crossing zero at 1.05 eV ($t\ \textgreater\ 1$ ps). Representative spectra of the latter two features are shown in Fig. 2\ \textbf{b} (the coherent artefact is not shown here because it is just an experimental artefact arising from the temporal overlap of the pump and probe pulses). The decay of the negative peak at 1.08 eV (Fig.\ 2 \textbf{c}) is fit well with a mono-exponential decay of the form $\frac{\Delta R(t)}{R} = Ae^{-t/\tau} + C$, in which $\tau$ is the lifetime and $C$ is a constant that accounts for the offset caused by the long-lasting broad feature. From the fit, the lifetime of this decay is $\sim0.2$ ps. The broad feature, although fit well enough by a constant on the picosecond timescale, eventually decays on the nanosecond timescale.

\begin{figure*}
\includegraphics[width=0.75\textwidth]{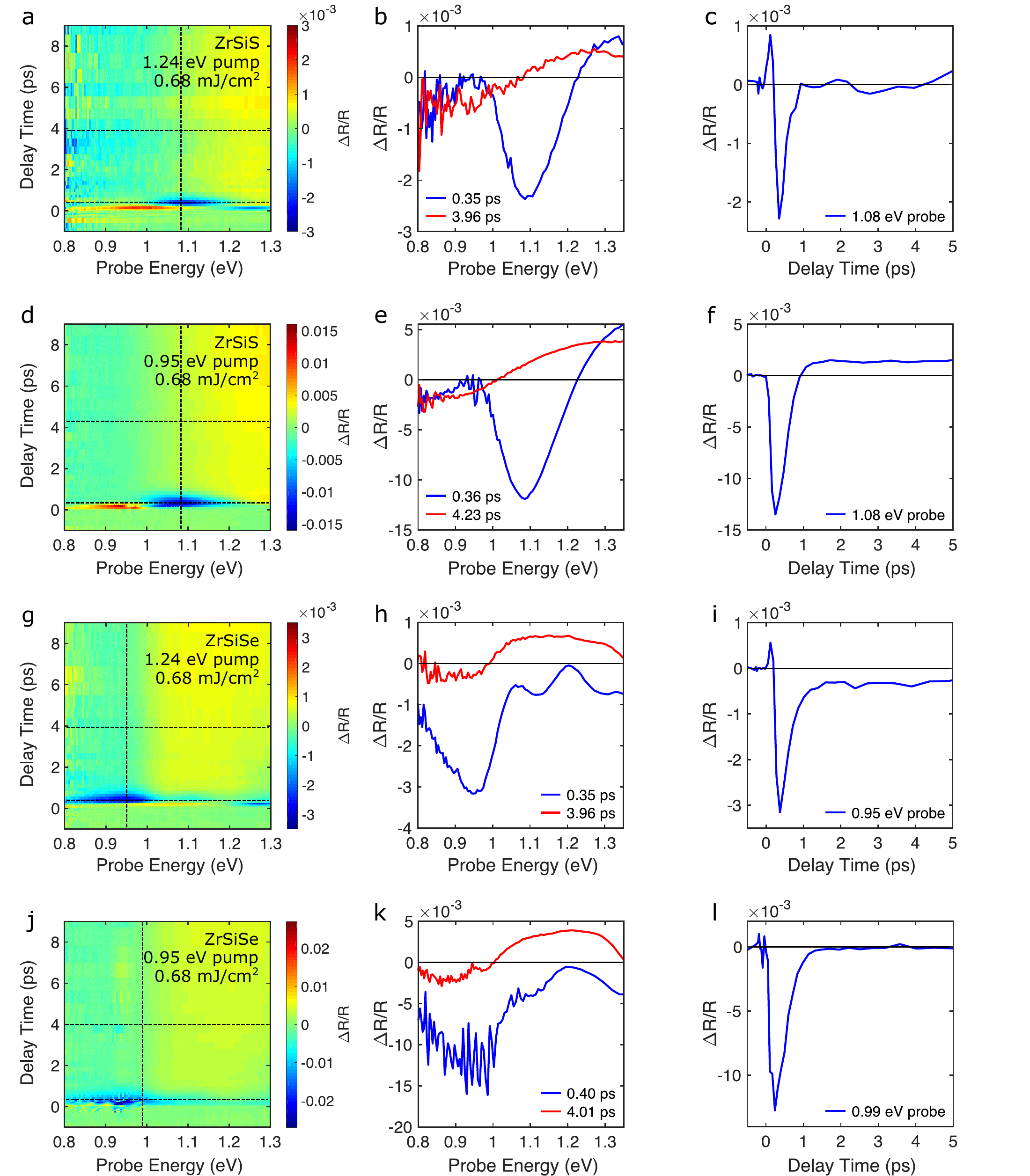}
\caption{\label{fig} \textbf{(a)}$-$\textbf{(c)} show the transient response of ZrSiS pumped at 1.24 eV: \textbf{(a)} tracks the transient reflectivity as a function of the probe energy and time delay between pump and probe pulses, \textbf{(b)} highlights the two main spectral features observed, and \textbf{(c)} shows the decay dynamics of the negative peak at 1.08 eV. The initial positive spike is the coherent artefact. \textbf{(d)}$-$\textbf{(f)} show the transient response of ZrSiS pumped at 0.95 eV: \textbf{(d)} tracks the transient reflectivity as a function of the probe energy and time delay between pump and probe pulses, \textbf{(e)} highlights the two main spectral features observed, and \textbf{(f)} shows the decay dynamics of the negative peak at 1.08 eV. \textbf{(g)} - \textbf{(i)} show the transient response of ZrSiSe pumped at 1.24 eV: \textbf{(g)} tracks the transient reflectivity as a function of the probe energy and time delay between pump and probe pulses, \textbf{(h)} highlights the two main spectral features observed, and \textbf{(i)} shows the decay dynamics of the negative peak at 0.95 eV. \textbf{(j)}$-$\textbf{(l)} show the transient response of ZrSiSe pumped at 0.95 eV: \textbf{(j)} tracks the transient reflectivity as a function of the probe energy and time delay between pump and probe pulses, \textbf{(k)} highlights the two main spectral features observed, and \textbf{(l)} shows the decay dynamics of the negative peak at 0.99 eV. 0.99 eV was chosen here in lieu of 0.95 eV because the dynamics at that wavelength are obscured by pump scatter. Vertical black lines in \textbf{(a)}, \textbf{(d)}, \textbf{(g)}, and \textbf{(j)} indicate the transients shown in \textbf{(c)}, \textbf{(f)}, \textbf{(i)}, and \textbf{(l)}, respectively, and horizontal black lines indicate the spectra shown in \textbf{(b)}, \textbf{(e)}, \textbf{(h)}, and \textbf{(k)}, respectively.}
\end{figure*}

Similar data for ZrSiS pumped at 0.95 eV are shown in Fig.\ 2 \textbf{d}-\textbf{f}. When excited with the lower energy pump pulse, the material's transient response is almost exactly identical to that of the higher-energy pump. There are a few differences, however: the coherent artefact shifts spectrally to lower energy, which is consistent with the lower energy pump pulse, the zero of the broad feature also shifts to slightly lower energy (1 eV instead of 1.05 eV), and the magnitude of the response \textemdash\ both the 1.08 eV peak and the broad feature \textemdash\ are almost an order of magnitude larger than when pumped at 1.24 eV. Like the 1.24 eV-pump experiments, the peak at 1.08 eV decays with a lifetime of $\sim0.2$ ps, and the broad feature lasts for nanoseconds. 

Moving on to ZrSiSe, we find spectral features almost identical to those observed in ZrSiS. The response of ZrSiSe to the 1.24 eV pump (Fig.\ 2 \textbf{g}-\textbf{i}) consists of a comparable set of three spectral features, occurring, however, at slightly lower energies than in ZrSiS. The negative peak centered at 0.95 eV (Fig.\ 2 \textbf{i}), which is analogous to the 1.08 eV peak in the ZrSiS measurements, decays with a $\sim0.3$ ps lifetime (slightly longer than the $\sim0.2$ ps lifetime in ZrSiS), and the broad feature, which crosses zero around 1 eV, lasts for nanoseconds. The only feature that does not shift in energy is the coherent artefact, which further reinforces that this is an experimental artefact. 

When pumped at 0.95 eV (Fig.\ 2 \textbf{j}-\textbf{l}), ZrSiSe responds exactly as it did when pumped at 1.24 eV, except that again the coherent artefact shifts to lower energies and the magnitude of the features are roughly an order of magnitude larger than when pumped with 1.24 eV photons. This is similar to the differences in the ZrSiS measurements. There is also significantly more scatter from the pump pulse in these data, which can obscure some of the dynamics; for clarity's sake, the transient shown in Fig.\ 2 \textbf{l} is at 0.99 eV rather than at the scatter-heavy minimum at 0.95 eV. 

The fluence dependence of these time constants was also investigated. Over an order of magnitude of fluences ranging from 0.17 mJ/cm$^{2}$ to 1.2 mJ/cm$^{2}$, the 0.2 ps lifetime of the 1.08 eV peak in ZrSiS and the 0.3 ps lifetime of the 0.95 eV peak in ZrSiSe remain essentially constant. The spectral profile of their features also remain constant over this range of fluences, and the magnitude increases linearly with fluence (SI). 

To understand the physical origin of these transient spectral features, we turn to the equilibrium reflectance spectra of the two materials, collected from the MIR to the visible (Fig.\ 3). The spectra for both materials consist of high reflectivity in the MIR (R \textgreater\ 0.8), a plasma reflection edge around 1 eV, and higher-energy interband transitions in the visible region. The plasma reflection edge appears at slightly lower energies in ZrSiSe than in ZrSiS. So does the onset of interband transitions, consistent with ZrSiSe’s slightly compressed band structure. These spectra match the equilibrium reflection spectra for these materials in Refs. \cite{Schill2017, Ebad2019} quite well. Drude-Lorentz fits for these spectra, each consisting of 1 Drude peak and 6 Lorentzians, are shown in Fig.\ 3, overlaid on the experimental spectra. We can see that our probe window, 0.8 eV \textendash\ 1.35 eV, straddles the plasma reflection edge. The Drude-Lorentz fits also confirm that the screened plasma frequencies, $\omega_p$, of the materials lie in this region: $\omega_p = 1.06$ eV for ZrSiS and $\omega_p = 1$ eV for ZrSiSe. These are very close in energy to the spectral features observed in the transient data (1.08 eV for ZrSiS, 0.95 eV for ZrSiSe): both the negative peak and the $\frac{\Delta R}{R}=0$\ part of the later broad feature are centered very near these plasma frequencies.

\begin{figure}
\includegraphics[width=0.4\textwidth]{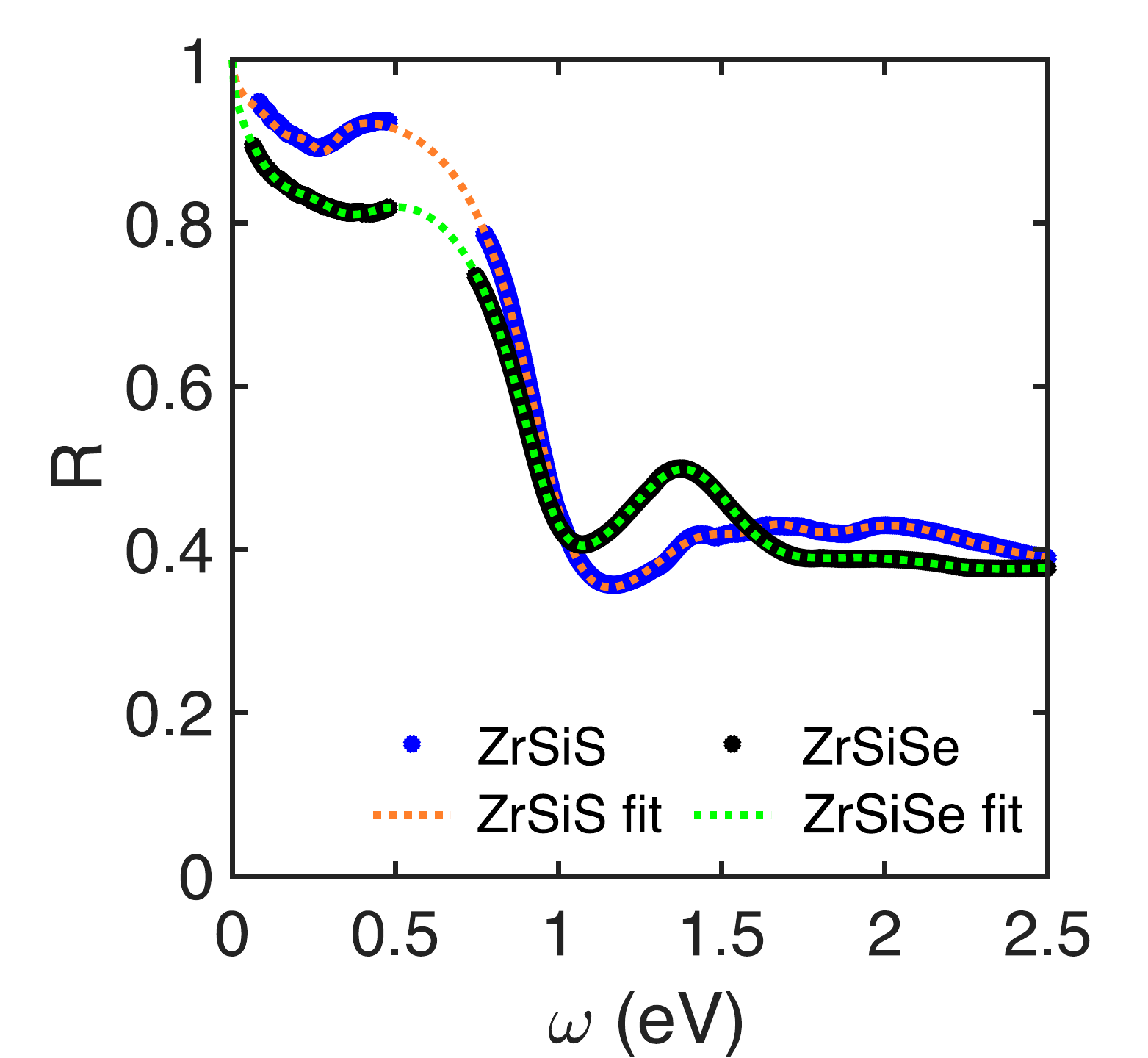}
\caption{\label{fig} Equilibrium reflectance spectra for ZrSiS (solid blue line) and ZrSiSe (solid black line). The Drude-Lorentz fits to the spectra (dashed red line for ZrSiS, dashed green line for ZrSiSe) are also presented; fit parameters can be found in the SI. The energy ranges of the reflectance spectrometers used for these measurements do not overlap, resulting in the gap in experimental data from 0.5 eV \textendash\ 0.75 eV.}
\end{figure}

With this in mind, we were able to envisage a simple transient Drude picture \cite{Giannetti2016} that accounts for the observed spectral features (Fig.\ 4, \textbf{a}, \textbf{d}). A decrease in the Drude plasma frequency, leaving all other Drude and Lorentz parameters fixed (Fig.\ 4, \textbf{d}), would lead to a decrease in the reflectance across the plasma reflection edge, which in turn results in a negative feature at that energy in the transient response (i.e. $\frac{R(\Delta\omega_p \textless 0) - R(\Delta\omega_p = 0)}{R(\Delta\omega_p = 0)}$); this matches the initial negative feature. In a similar fashion, increasing the Drude scattering rate (again leaving all the other fit parameters fixed, now including the Drude plasma frequency, too) leads to a decreased (increased) reflectance below (above) $\omega_p$, similar to that of the long-lived broad features.

To check this model against our data, the transient reflectivity spectra at each pump-probe time delay were converted to reflectance spectra and fit with the Drude-Lorentz models discussed above \cite{Novelli2012}. Leaving all parameters fixed, save the Drude plasma frequency and scattering rate, these fits give the time evolution of these parameters (Fig.\ 4, \textbf{b}, \textbf{c}, \textbf{e}, \textbf{f}). In both materials, at all pump fluences, the drude plasma frequency initially decreases, recovering after a few hundred femtoseconds, while the scattering rate increases after roughly 1 ps, remaining elevated for picoseconds. Higher pump fluences result in greater decreases of the plasma frequency and greater increases in the scattering rate.

\begin{figure*}
\includegraphics[width=1\textwidth]{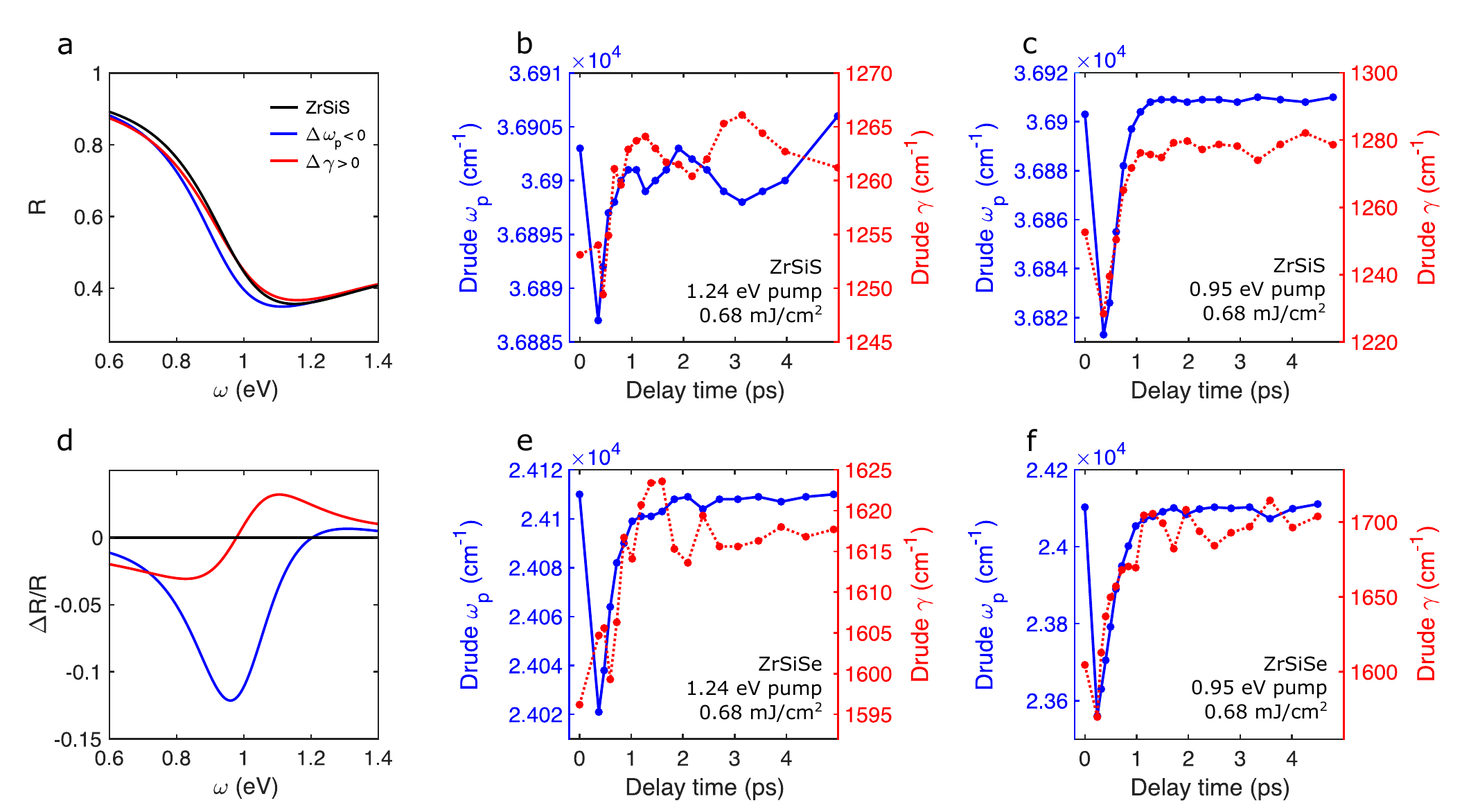}
\caption{\label{fig} \textbf{(a)} The impact of decreasing the Drude plasma frequency and, separately, increasing the Drude scattering rate on the model reflectance of ZrSiS. The black line is the equilibrium reflectance. The way these changes to the Drude response appear in the transient reflectivity spectrum are shown in \textbf{(d)}, where the blue spectrum corresponds to a decreased plasma frequency and the red spectrum corresponds to an increased scattering rate. \textbf{(b)}, \textbf{(c)}, \textbf{(e)}, and \textbf{(f)} show the evolution of these two Drude parameters extracted from the Drude-Lorentz fits of the transient data for ZrSiS pumped at 1.24 eV, ZrSiS pumped at 0.95 eV, ZrSiSe pumped at 1.24 eV, and ZrSiSe pumped at 0.95 eV, respectively. Details of the fits are discussed in the Results and Experimental sections.}
\end{figure*}

Comparing the responses to the two different pump energies, 1.24 eV and 0.95 eV, at time delays for which the pulses are no longer overlapped there is no difference in the dynamics or the spectral features. From this we can say that excited carriers end up in the same region of the CB within $\sim100$ fs. With the time resolution of our experiment, we cannot tell whether the two pump energies only induce transitions between the nodal line bands, or whether the higher energy pump excites carriers to a higher band, and they quickly relax to the nodal-line bands, or a combination of both. It is assumed that carrier thermalization also happens within the time resolution. The differences in excitation density at the two pump energies (discussed in the SI) likely contribute to the differences in the magnitude of the response. 

This initial excitation of carriers from the nodal line bands near $E_F$ to states further in energy from $E_F$ is what leads to the initial decrease of the Drude plasma frequency in our model. As the excited carriers relax back towards $E_F$, the concentration of \textquotedblleft free carriers\textquotedblright\ increases back towards its equilibrium value, resulting in the recovery of the equilibrium value of $\omega_p$. As the pump fluence increases, a greater number of carriers are being removed from $E_F$, which results in a larger initial decrease in $\omega_p$. These carriers lose their excess energy by emitting phonons; the production of these phonons elevates the lattice temperature relative to the ambient environment. In the Drude model, this is equivalent to increasing the scattering rate of the phononic bath. Concomitant with the recovery of $\omega_p$, $\gamma$ increases above it's equilibrium value, remaining at a relatively constant elevated value for picoseconds. In some of the fits $\gamma$ deceases initially, similar to $\omega_p$; likely the contributions of each parameter to the fit are not perfectly separable, but overall this model is supported by our data. 

The 0.2 ps lifetime of the 1.08 eV peak in ZrSiS is consistent with the 0.19 ps and 0.2 \textendash\ 0.4 ps lifetimes observed at 1.55 eV and in the MIR. From our transient Drude fits, we can assign the process behind this lifetime to the cooling of excited hot carriers as they return to the vicinity of the Fermi energy. This is seen in our experiments as the recovery of the equilibrium Drude plasma frequency. The lattice heating observed here likely accounts for the 100 ps lifetime in the 1.55 eV study. The MIR study saw no such signature of lattice heating because the MIR probe signal was dominated by the effect of phase-space filling. The slight difference in lifetimes between ZrSiS and ZrSiSe (0.2 ps vs 0.3 ps) is likely due to the slightly higher phonon energies in ZrSiS compared to those in ZrSiSe \cite{Salmankurt2016, Singha2018, Zhou2017}. Higher energy phonons would allow excited carriers to lose more energy per phonon emission, and if the phonon emission rates are similar the excited carriers in the material with the higher energy phonons should cool more rapidly. The nodal-line bands in ZrSiS are also slightly steeper than those in ZrSiSe \cite{Neupane2016, Fu2019, Hosen2017}; this might also expedite the cooling process. Lastly, the change could also originate from different carrier concentrations in these two different batches of materials.

Details of the process through which the hot carriers cool are still elusive, but, as posited earlier, these data suggest that it occurs simply through the coupling of the out-of-equilibrium electronic population with the phonon bath of the lattice, such that phonons are emitted as the excited carriers lose energy. The lattice heating observed \textemdash\ a result of such a transiently excited phonon population \textemdash\ supports this electron-phonon coupling cooling process. The linear band structures of these materials appear as if they should facilitate Auger recombination, in which the energy emitted from the relaxation of one carrier is immediately absorbed by another carrier, prolonging the excited state. There is, however, no evidence to support the occurrence of this process in these experiments. Auger recombination is a multi-carrier process and would result in a lifetime with a strong pump fluence dependence, which we do not observe. This result supports a recent theoretical study that predicts that Auger recombination is actually suppressed in TSMs, for relativistic reasons \cite{Afanasiev2019}. The long-lasting lattice heating \textemdash\ observed in the enhancement of the Drude scattering rate, and the result of a hot phonon population \textemdash\ measures the timescale on which heat is transported out of the laser spot. 

Transient reflectivity measurements in the NIR were performed on the NLSMs ZrSiS and ZrSiSe. Both materials exhibit similar ultrafast responses characterized by two spectrally and kinetically distinct features. The first is a negative peak occurring immediately after the coherent artefact and decaying with a 0.2 ps lifetime in ZrSiS and a 0.3 ps lifetime in ZrSiSe, and the second is a broader negative-to-positive feature that grows in after 1 ps and decays on the nanosecond timescale. The hundreds-of-femtoseconds lifetime is fairly independent of pump fluence, over almost an order of magnitude from 0.17 mJ/cm$^{2}$ to 1.2 mJ/cm$^{2}$. Drawing from Drude-Lorentz fits to the equilibrium reflectance of the two materials, we were able to directly describe the transient response with an increase and subsequent return to equilibrium of the Drude plasma frequency, followed by an increase in the Drude scattering rate. This paints a physical picture in which carriers are removed from the vicinity of the Fermi energy and then relax back, and the energy lost by the relaxing carriers heats the lattice.
Therefore, electronic relaxation within the nodal line structure obeys the conventional picture in which hot carriers cool through phonon emission. The rapid carrier cooling in the nodal line bands continues to make these materials promising candidates for incorporation in high-frequency optoelectronic devices.

\section{Experimental}

Single crystals of ZrSiS and ZrSiSe were grown by chemical vapor transport. Zirconium sponge (Sigma Aldrich, 99\%), silicon pieces (Sigma Aldrich, 99.95\%), and sulfur (Alfa Aesar, 99.5\%) or purified selenium crystals, for ZrSiS and ZrSiSe respectively, were stoichiometrically mixed and small amounts of iodine (Alfa Aesar, ultra dry 99.999\%) added. The purified selenium crystals were obtained by chemical vapor transport using selenium pellets (Sigma Aldrich, 99.99\%). The elements were sealed in a quartz ampoule \textit{in vacuo}. The tube was placed in a tube furnace and heated to 1100$^{\circ}$C in 6h and held for 7 days. After this time the furnace was cooled to room temperature within 6h. Representative crystals were ground to a powder and analyzed using a STOE STADI P x-ray diffractometer (Mo K$_{\alpha1}$ radiation, Ge monochromator).

For the pump-probe experiments, the output from a 1 kHz regeneratively-amplified Ti:sapphire laser (Coherent Libra, Santa Clara, California), which produces pulses centered at 1.55 eV with a pulse duration of $\sim45$ fs, was split with a 50:50 beamsplitter to generate the pump and probe pulses. Pump pulses were generated in a commercial optical parametric amplifier (OPerA Solo, Vilnius, Lithuania), producing $\sim60$ fs narrowband pulses centered at 0.95 eV and 1.24 eV. The pump pulse intensity was controlled with a neutral-density filter, and was varied from 0.17 mJ/cm$^{2}$ to 1.2 mJ/cm$^{2}$. Probe pulses were generated by focusing the probe portion of the 1.55 eV light in a sapphire crystal, producing NIR \textquotedblleft white light\textquotedblright\ from 0.8 eV to 1.35 eV; an iris controls the shape of the beam, and a $\lambda/2$-waveplate and polarizer control the polarization and intensity of the light before it is used to generate the broadband pulse. The measurements themselves were performed in a commercial pump-probe setup (Ultrafast System Helios, Sarasota, Florida), in the reflection geometry. Pump and probe pulses arrive at the sample at near-normal incidence. A delay line in the probe arm, capable of up to 7 ns delay, controls the time delay between when the pump and probe pulses arrive at the sample. The reflected probe light is focused into a fiber optic, leading to a detector. The transient reflectivity is presented as $\frac{\Delta R}{R}(\omega,t) = \frac{(R(\omega,t)_{pump\ on} - R(\omega, t)_{pump\ off})}{R(\omega,t)_{pump\ off}}$; the pump arm contains a 500 Hz chopper, to provide the `pump on' and `pump off' conditions. 
 
ZrSiSe is air-sensitive, so transient measurements of both materials were performed with the single crystals housed in an air-tight chamber prepared inside an argon-filled glovebox. Prior to measuring, crystals were cleaved along the $a-b$ plane to present a fresh surface. All measurements were performed at room temperature. 

The raw data are background-corrected by averaging the transient response over the negative delay times (i.e. the weak probe pulse reaches the sample before the stronger pump pulse) and subtracting that average from the entire data set. The broadband probe pulse accumulates chirp as it is generated in the sapphire crystal, making the lower-energy response appear at later delay times than at higher energies; this is corrected using the Helios proprietary software Surface Xplorer. Points along the chirped coherent artefact are fit with a polynomial function (delay time as a function of probe energy), and subtracted such that all energies along the polynomial occur at the same delay time. 

Equilibrium reflectance measurements were performed from the MIR to the visible. MIR reflectance spectra were recorded with an FTIR microscope (Nicolet iN10 MX), using gold as a reference, and NIR and visible reflectance spectra were recorded with a UV/Vis spectrometer (Cary 6000i) outfitted with an integrating sphere, using a silver reference. The reflectance spectra were fit with Drude-Lorentz models using the optical fitting program RefFit \cite{Kuzmenko2005}; the Drude-Lorentz parameters from these fits can be found in the Supplemental Information (SI). Transient Drude-Lornetz fitting makes use of the reflectance of the materials as a function of the delay time, $R(\omega,t)$, as opposed to $\frac{\Delta R}{R}(\omega,t)$; this conversion is done by $R(\omega,t) = R_{eq}(\omega)(1 + \frac{\Delta R}{R}(\omega,t))$, where $R_{eq}(\omega)$ is the equilibrium, or steady-state, reflectance of the material.

\begin{acknowledgement}

This work was supported by NSF through the Princeton Center for Complex Materials, a Materials Research Science and Engineering Center DMR-1420541, by Princeton University through the Princeton Catalysis Initiative, and by the Gordon and Betty Moore Foundation through Grant GBMF9064 to L.M.S. G.D.S. is a CIFAR Fellow in the Bio-Inspired Energy Program. D.F. was supported by the European Commission through the European Research Council (ERC), Project INCEPT, Grant 677488.

\end{acknowledgement}

\begin{suppinfo}

Calculation of excitation densities, permittivities and penetration depths from equilibrium relflectance fits, additional transient data, linearity of the transient response, transient lifetimes, additional transient Drude-Lorentz fits, and parameters from the equiilibrium Drude-Lorentz fits. 

\end{suppinfo}


\providecommand{\noopsort}[1]{}\providecommand{\singleletter}[1]{#1}%
\providecommand{\latin}[1]{#1}
\makeatletter
\providecommand{\doi}
  {\begingroup\let\do\@makeother\dospecials
  \catcode`\{=1 \catcode`\}=2 \doi@aux}
\providecommand{\doi@aux}[1]{\endgroup\texttt{#1}}
\makeatother
\providecommand*\mcitethebibliography{\thebibliography}
\csname @ifundefined\endcsname{endmcitethebibliography}
  {\let\endmcitethebibliography\endthebibliography}{}

\end{document}